\documentstyle[12pt]{article}

\def\seCtion#1{\section{#1} \setcounter{equation}{0}}
\renewcommand\theequation{\ifnum\value{section}>0{\thesection.
\arabic{equation}}\fi}
\newcommand{\be}{\begin{equation}}
\newcommand{\ee}{\end{equation}}
\newcommand{\bea}{\begin{eqnarray}}
\newcommand{\eea}{\end{eqnarray}}
\newcommand{\nn}{\nonumber}
\newcommand{\bi}{\bibitem}
\newcommand{\la}{\label}

\begin{document}

\pagestyle{empty}
\begin{flushright}
FIUN-GCP-07/1
\end{flushright}

\begin{center}
{\Large \bf Effective model for particle mass generation}\\

\vspace{0.3cm} {\bf C. Quimbay \footnote{cjquimbayh@unal.edu.co,
associate researcher of CIF, Bogot\'a, Colombia.} and J. Morales
\footnote{jmoralesa@unal.edu.co,
associate researcher of CIF, Bogot\'a, Colombia.}}\\
{\it{Departamento de F\'{\i}sica}, Universidad Nacional de
Colombia\\
Ciudad Universitaria, Bogot\'a, Colombia}\\
\vspace{0.5cm} May 13, 2008
\end{center}

\vspace{0.8cm}

\begin{abstract}
We present an effective model for particle mass generation in
which we extract generic features of the Higgs mechanism that do
not depend on its interpretation in terms of a Higgs field. In
this model the physical vacuum is assumed as a medium at zero
temperature which is formed by virtual fermions and antifermions
interacting among themselves through the intermediate gauge bosons
of the standard model without Higgs sector. As a consequence the
fermions acquire their masses from theirs interactions with the
vacuum and the gauge bosons from the charge fluctuations of the
vacuum. This effective model is completely consistent with the
physical mass spectrum, in such a way that the left-handed
neutrinos are massive. The masses of the electroweak gauge bosons
are properly predicted in terms of the experimental fermions
masses and the running coupling constants of the strong,
electromagnetic and weak interactions.

\end{abstract}

\newpage

\pagestyle{plain}


\seCtion{Introduction}

The Higgs mechanism is the currently accepted procedure to
generate the masses of electrically charged fermions and
electroweak bosons in particle physics \cite{djouadi}. The
implementation of this mechanism requires the existence of a
sector of scalar fields which includes, in the Lagrangian density
of the model, a Higgs potential and Yukawa terms. In the Minimal
Standard Model (MSM), the Higgs field is a doublet in the
$SU(2)_L$ space carrying non-zero hypercharge, and a singlet in
the $SU(3)_C$ space of color. The Higgs mechanism is based on the
fact that the neutral component of the Higgs field doublet
spontaneously acquires a non-vanishing vacuum expectation value.
Since the vacuum expectation value of the Higgs field is different
from zero, the Higgs field vacuum can be interpreted as a medium
with a net weak charge. In this way the $SU(3)_C \times SU(2)_L
\times U(1)_Y$ symmetry is spontaneously broken into the $SU(3)_C
\times U(1)_{em}$ symmetry \cite{weinberg}.

As a consequence of the MSM Higgs mechanism, the electroweak gauge
bosons acquire their masses, in such a way that the masses depend
of the vacuum expectation value of the Higgs field, which is a
free parameter in the MSM. This parameter can be fixed by means of
the calculation of the muon decay, at tree level, using the Fermi
effective coupling constant. Simultaneously, Yukawa couplings
between the Higgs field and fermion fields lead to the generation
of masses for the electrically charged fermions, that depend of
the Yukawa coupling constants, which also are free parameters in
the MSM. These constants can be fixed by mean of the experimental
values of the fermionic masses. The above mechanism implies the
existence of a neutral Higgs boson in the physical spectrum, whose
mass is a free parameter in the model. Because there only exist
left-handed neutrinos in the Lagrangian density of the MSM, after
the spontaneously electroweak symmetry breaking neutrinos remain
massless.

In the current picture of the Higgs mechanism \cite{djouadi} the
masses of the MSM particles spectrum are generated through the
interactions among the electroweak gauge bosons and the
electrically charged fermions with the weakly charged Higgs field
vacuum. However, there are some physical aspects in this picture
of mass generation that are not completely satisfactory and that
we summarize in the following questions: What is the possible
description of the interactions among the fermions and the
electroweak bosons with the Higgs field vacuum? How is it possible
to show in a fundamental way that the particle masses are
generated by these interactions? Why the origin of the particle
masses is only related to the weak interaction? Why the
most-intense interactions (strong and electromagnetic) are not
related to the mass generation mechanism? Why are there no
interactions among the weakly charged left-handed neutrinos and
the weakly charged Higgs field vacuum? Why the left-handed
neutrinos are massless if they have a weak charge?

All the above questions might have a trivial answer if we only
look at things through the current picture of the Higgs mechanism.
However, we are interested in to research the possible physics
behind the Higgs mechanism. In this way we propose an effective
model for particle mass generation in which fermions acquire their
masses from theirs interactions with the physical vacuum and gauge
bosons from the charge fluctuations of the vacuum.

The physical vacuum is the state of lowest energy of all gauge
bosons and fermion fields \cite{schwinger1}. As it is well known
from the covariant formulation of the quantum field theory
\cite{schwinger2} the state of lowest energy of the gauge bosons
and matter fields has infinite energy. This physical vacuum is a
rich medium where there are processes involving virtual massless
particles and virtual massless antiparticles with unlimited
energy. The physical vacuum is then assumed as a virtual medium at
zero temperature which is formed by massless fermions and
antifermions interacting among themselves exchanging massless
gauge bosons.

The fundamental model describing the dynamics of the physical
vacuum is the Standard Model without the Higgs Sector (SMWHS),
which is based in the $SU(3)_C \times SU(2)_L \times U(1)_Y$ gauge
symmetry group. We assume that each fermion in the physical vacuum
has associated a chemical potential which describes the excess of
virtual antifermions over virtual fermions. Then there are twelve
fermionic chemical potentials $\mu_{f}$ associated with the six
leptons and the six quarks implying an antimatter-matter asymmetry
in the physical vacuum. Hence, the physical vacuum is considered
as a virtual medium having antimatter finite density. This
antimatter-matter asymmetry of the physical vacuum is related with
CP violation by the electroweak interactions. Naturally, the
mentioned asymmetry has an inverse sign respect to the one of the
matter-antimatter asymmetry of the Universe. The existence of
fermionic chemical potentials in the physical vacuum does not
implies that this vacuum itself carries net charges. This can be
understood in a similar way as the existence of the maximal
matter-antimater asymmetry of the Universe does not mean that
baryonic matter carries net charges.

The masses of fermions are obtained starting from their
self-energies, which give account of the fundamental interactions
of massless fermions with the physical vacuum \cite{schwinger1}.
While quark masses are generated by strong, electromagnetic and
weak interactions, the electrically charged lepton masses are only
generated by electromagnetic and weak interactions, and the masses
of neutrinos are generated from the weak interactions. On the
other hand, gauge bosons masses are obtained from the charge
fluctuations of the physical vacuum, which are described by the
vacuum polarization tensors \cite{schwinger1}.

We use the following general procedure to calculate the particle
masses: Initially we write the one-loop self-energies and the
one-loop polarization tensors at finite density and finite
temperature; next we calculate the dispersion relations by
obtaining the poles of the fermion and gauge boson propagators;
from these dispersion relations we find the fermion and gauge
boson effective masses at finite density and finite temperature;
finally, we identify these effective masses at zero temperature
with the physical masses. This identification can be performed
because the virtual medium at zero temperature is representing the
physical vacuum.

From a different perspective other works have intended to show
that the inertial reaction force, appearing when a macroscopic
body is accelerated by an external agent, is originated in a
reaction by the physical vacuum that opposes the accelerating
action \cite{arueda1}. These works involved only the
electromagnetic quantum vacuum and have been able to yield the
$F=ma$ expression as well as its relativistic generalization. An
expression for the contribution by the electromagnetic quantum
vacuum to the inertial mass of a macroscopic object has been found
and this has been extended to the gravitational case. Originally
these works have used a semiclassical approach \cite{arueda1}
which has been easily extended to a quantum version
\cite{arueda2}.

We find that the fermion and gauge boson masses are functions of
the vacuum fermionic chemical potentials $\mu_{f}$, which are
fixed using the experimental fermion masses. Then using the values
of the all fermionic chemical potentials obtained we calculate the
masses of the electroweak gauge bosons obtaining an agreement with
their experimental values. The main result of this effective model
for particle mass generation is that the left-handed neutrinos are
massive because they have weak charge. The weak interaction among
the massless neutrinos and the physical vacuum is the source of
the neutrino masses.

Before considering the real case of the physical vacuum described
by the SMWHS, in section 2 we first show how to obtain the gauge
invariant masses of a fermion and a gauge boson for the case in
which the dynamics of the vacuum is described by a non-abelian
gauge theory. In section 3 we consider the SMWHS as the model
which describes the dynamics of the physical vacuum and we obtain
the fermions (quarks and leptons) and the electroweak gauge bosons
($W^{\pm}$ and $Z^0$) masses. We obtain consistently the masses of
the electroweak gauge bosons in terms of the masses of the
fermions and the running coupling constants of the three
fundamental interactions. In section 4 we focus our interest in to
find a restriction about the possible number of families, a
prediction of the top quark mass, a highest value for the summing
of the square of the neutrino masses. Our conclusions are
summarized in section 5.


\seCtion{Non-abelian gauge theory case}

In this section we first consider a more simple case in which the
dynamics of the vacuum is described by a non-abelian gauge theory,
and in this context we calculate the fermion and gauge boson masses.
The vacuum is assumed to be a quantum medium at zero temperature
constituted by virtual massless fermions and antifermions
interacting among themselves through massless non-abelian gauge
bosons. We also assume that there exist an excess of virtual
antifermions over virtual fermions in the vacuum. This
antimatter-matter asymmetry of the vacuum is described by
non-vanishing fermionic chemical potentials $\mu_{f_i}$, where $f_i$
represents the different fermion species. In this section, for
simplicity we will take $\mu_{f_1}= \mu_{f_2}= \ldots = \mu_f$.

The non-abelian gauge theory describing the dynamics of the vacuum
is given by the following Lagrangian density \cite{weldon}:

\be {\cal L}=-\frac{1}{4} F_{A}^{\mu \nu} F_{\mu \nu}^{A} +
\bar{\psi}_m \gamma^{\mu} \left( \delta_{mn} i \partial_{\mu} + g
L_{mn}^{A} A_{\mu}^{A} \right) \psi_n , \la{lag} \ee where $A$ runs
over the generators of the group and $m,n$ over the states of the
fermion representation. The covariant derivative is $\mbox{D}_\mu =
\delta_\mu + igT_A A_{\mu}^A$, being $T_A$ the generators of the
$SU(N)$ gauge group and $g$ the gauge coupling constant. The
representation matrices $L_{mn}^{A}$ are normalized by $Tr(L^A L^B)
= T(R) \delta^{AB}$ where $T(R)$ is the index of the representation.
In the calculation of the fermionic self-energy appears $(L^A
L^A)_{mn} = C(R) \delta_{mn}$, where $C(R)$ is the quadratic Casimir
invariant of the representation \cite{weldon}.

At finite temperature and density, Feynman rules for vertices are
the same as those at $T=0$ and $\mu_f=0$, while the propagators in
the Feynman gauge for massless gauge bosons $D_{\mu \nu}(p)$,
massless scalars $D(p)$ and massless fermions $S(p)$ are
\cite{kobes}: \bea D_{\mu \nu}(p) &=& -g_{\mu \nu} \left[
\frac{1}{p^2+i\epsilon} -i
\Gamma_b(p) \right],  \la{bp} \\
D(p) &=& \frac{1}{p^2+i\epsilon}-i{\Gamma}_b(p),  \la{ep} \\
S(p) &=& \frac{p{\hspace{-1.9mm}\slash}}{p^2+i \epsilon}+
i p{\hspace{-1.9mm} \slash}{\Gamma}_f(p),  \la{fp}
\eea
where $p$ is the particle four-momentum and the plasma temperature $T$
is introduced through the functions $\Gamma_b(p)$ and $\Gamma_f(p)$,
which are given by
\bea
\Gamma_b (p)= 2\pi \delta(p^2)n_b (p),  \la{db} \\
\Gamma_f (p)= 2\pi \delta(p^2)n_f (p),  \la{df}
\eea
with
\bea
n_b (p) &=& \frac{1}{e^{|p\cdot u|/T}-1}, \la{nb}\\
n_f(p) &=& \theta(p\cdot u)n_{f}^{-}(p)+\theta(-p\cdot
u)n_{f}^{+}(p), \la{nf} \eea  being $n_b(p)$ the Bose-Einstein
distribution function. The Fermi-Dirac distribution functions for
fermions $n_{f}^{-}(p)$ and for anti-fermions $n_{f}^{+}(p)$ are:
\bea n_{f}^{\mp}(p)= \frac{1}{e^{(p\cdot u \mp \mu_f)/T}+1}. \eea In
the distribution functions $(\ref{nb})$ and $(\ref{nf})$,
$u^{\alpha}$ is the four-velocity of the center-of-mass frame of the
dense plasma, with $u^\alpha u_\alpha =1$.

\subsection{Self-energy and fermion mass}

We first consider the propagation of a massless fermion in a medium
at finite density and finite temperature. The finite density of the
medium is associated with the fact that it has more antifermions
than fermions. The fermion mass is calculated by following the
general procedure describe in the introduction.

For a non-abelian gauge theory with parity and chirality
conservation, the real part of the self-energy for a massless
fermion is written as: \be \mbox{Re}\,\Sigma^{\prime}(K)=-
aK{\hspace{-3.1mm}\slash}-b u{\hspace{-2.1mm} \slash},  \la{tse} \ee
$a$ and $b$ being the Lorentz-invariant functions and $K^{\alpha}$
the fermion momentum. These functions depend on the Lorentz scalars
$\omega$ and $k$ defined as {\hspace {0.1 cm}} $\omega\equiv(K\cdot
u)$ and $k\equiv[(K\cdot u)^2-K^2]^{1/2}$. Taking for convenience
$u^\alpha=(1,0,0,0)$ we have $K^2 =\omega^2-k^2$ and then, $\omega$
and $k$ can be interpreted as the energy and three-momentum,
respectively. Beginning with $(\ref{tse})$ it is possible to write:
\bea a(\omega,k) &=& \frac{1}{4k^2} \left[
Tr(K{\hspace{-3.1mm}\slash}\, \mbox{Re}\,\Sigma^{\prime})- \omega
Tr(u{\hspace{-2.1mm}
\slash}\,\mbox{Re}\,\Sigma^{\prime}) \right],  \la{lifa} \\
b(\omega,k) &=& \frac{1}{4k^2} \left[ (\omega^2-k^2)
Tr(u{\hspace{-2.1mm}\slash}\,\mbox{Re}\,\Sigma^{\prime})-
\omega Tr(K{\hspace{-3.1mm}\slash}\,\mbox{Re}\,\Sigma^{\prime})
\right].  \la{lifb}
\eea

The fermion propagator, including only mass corrections, is given by
\cite{weldon1} \be S(p)=\frac
1{K{\hspace{-3.1mm}\slash}-\mbox{Re}\,\Sigma ^{\prime }(K)}=
\frac{1}{r}\frac{\gamma^0 \omega n - \gamma_i k^i}{n^2 \omega^2 -
k^2}, \la{pft} \ee where $n = 1 + b(\omega,k)/r\omega$ and $ r = 1 +
a(\omega,k)$. The propagator poles can be found when: \be \left[ 1 +
\frac{b(w,k)}{w(1 + a(w,k))} \right]^2 w^2 - k^2 = 0.  \la{fdr0} \ee
We observe in $(\ref{fdr0})$ that $n$ plays a role similar to that
of the index of refraction in optics. To solve the equation
$(\ref{fdr0})$, $a(\omega,k)$ and $b(\omega,k)$ are first calculated
from the relations $(\ref{lifa})$ and$(\ref{lifb})$ in terms of the
real part of the fermionic self-energy. The contribution to the
fermionic self-energy from the one-loop diagram, which can be
constructed in this theory, is given by \be \Sigma (K)=ig^2C(R)\int
\frac{d^4p}{(2\pi )^4}D_{\mu \nu }(p) {\gamma }^\mu S(p+K){\gamma
}^\nu ,  \la{fse} \ee where $g$ is the interaction coupling constant
and $C(R)$ is the quadratic Casimir invariant of the representation.
For the fundamental representation of SU(N), $C(R)=(N^2-1)/2N$
\cite{weldon11}. We have that $C(R)=1$ for the U(1) gauge symmetry
group, $C(R)=1/4$ for SU(2) and $C(R)=4/3$ for SU(3).

Substituting $(\ref{bp})$ and $(\ref{fp})$ into $(\ref{fse})$, the
fermionic self-energy can be written as
$\Sigma(K)=\Sigma(0)+\Sigma^{\prime}(K)$, where $\Sigma(0)$ is the
zero-density and zero-temperature contribution and
$\Sigma^{\prime}(K)$ is the finite-density and finite-temperature
contribution. It is easy to see that: \bea \Sigma(0)=-ig^2C(R) \int
\frac{d^4 p}{(2\pi)^4} \frac{g_{\mu \nu}}{p^2} \gamma^{\mu}
\frac{p{\hspace{-1.9mm}\slash}+ K{\hspace{-3.1mm}\slash}} {(p+K)^2}
\gamma^{\nu} \eea and \bea \Sigma^{\prime}(K)=2g^2 C(R) \int
\frac{d^4 p}{(2\pi)^4} (p{\hspace{-1.9mm}\slash}+
K{\hspace{-3.1mm}\slash}) \left[
\frac{\Gamma_b(p)}{(p+K)^2}-\frac{\Gamma_f(p+K)}{p^2}+i \Gamma_b(p)
\Gamma_f(p) \right]. \eea Keeping only the real part
$(\mbox{Re}\,\Sigma^{\prime}(K))$ of the finite-density and
finite-temperature contribution, we obtain: \be
\mbox{Re}\,\Sigma^{\prime}(K)=2g^2C(R) \int \frac{d^4 p}{(2\pi)^4}
\left[(p{\hspace{-1.9mm}\slash}+ K{\hspace{-3.1mm}\slash})
\Gamma_b(p)+ p{\hspace{-1.9mm}\slash} \Gamma_f(p) \right]
\frac{1}{(p+K)^2}. \la{rse} \ee If we multiply $(\ref{rse})$ by
either $K{\hspace{-3.1mm}\slash}$ or $u{\hspace{-2.1mm}\slash}$,
take the trace and perform the integrations over $p_0$ and the two
angular variables, the functions $(\ref{lifa})$ and $(\ref{lifb})$
can be written in the notation given in \cite{quimbay} as: \bea
a(\omega,k)=g^2C(R)A(w,k,\mu_f), \la{aF0} \\
b(\omega,k)=g^2C(R)B(w,k,\mu_f), \la{bF0}
\eea
where the integrals over the modulus of the three-momentum
$p= \vert \vec{p} \vert$, $A(\omega,k,\mu_f)$ and $B(\omega,k,\mu_f)$,
are:

\bea A(\omega,k,\mu_f) &=&
\frac{1}{k^2}\int^\infty_0\frac{dp}{8\pi^2} \left[ 2p-\frac{\omega
p}{k} \log \left( \frac{\omega+k}{\omega-k} \right) \right]
\left[2n_b(p)+n_f^-(p)+n_f^+(p) \right], \la{Alead}
\nn \\
\\
B(\omega,k,\mu_f) &=& \frac{1}{k^2}  \nn \\
&\times& \int^\infty_0\frac{dp}{8\pi^2} \left[
\frac{p(\omega^2-k^2)}{k} \log \left( \frac{\omega+k}{\omega-k}
\right) -2\omega p \right]
\left[2n_b(p)+n_f^-(p)+n_f^+(p)\right]. \la{Blead} \nn \\
\eea The integrals $(\ref{Alead})$ and $(\ref{Blead})$ have been
obtained in the high density approximation, ${\it i. e.}$ $\mu_f \gg
k$ and $\mu_f \gg \omega$, and keeping the leading terms in
temperature and chemical potential \cite{morales}. Evaluating these
integrals we obtain that $a(\omega,k)$ and $b(\omega,k)$ are given
by: \bea a(\omega,k) &=& \frac{M_F^2}{k^2} \left[
1-\frac{\omega}{2k} \log
\frac{\omega+k}{\omega-k}\right], \la{a1} \\
b(\omega,k) &=& \frac{M_F^2}{k^2} \left[ \frac{\omega^2-k^2}{2k}
\log \frac{\omega+k}{\omega-k}-\omega \right], \la{b1} \eea where
the fermion effective mass $M_F$ is: \be M_F^2(T, \mu_f)=\frac{g^2
C(R)}{8} \left( T^2+\frac{\mu_f^2}{\pi^2} \right). \la{me} \ee The
value of $M_F$ given by $(\ref{me})$ is in agreement with
\cite{kajantie}-\cite{lebellac}. We are interested in the effective
mass at $T=0$, which corresponds precisely to the case in which the
vacuum is described by a virtual medium at zero temperature. For
this case: \be M_F^2 (0, \mu_F)= M_F^2=\frac{g^2 C(R)}{8}
\frac{\mu_f^2}{\pi^2}. \la{mef} \ee Substituting $(\ref{a1})$ and
$(\ref{b1})$ into $(\ref{fdr0})$, we obtain for the limit $k \ll
M_F$ that: \be \omega^2(k) = M_F^2 \left[ 1 + \frac{2}{3}
\frac{k}{M_F} + \frac{5}{9} \frac{k^2}{M_F^2} + \dots \right]
\la{dr1} \ee This dispersion relation is gauge invariant as the
calculation has been done at leading order in temperature and
chemical potential \cite{morales}.

It is well known that the relativistic energy in the vacuum for a
massive fermion at rest is $\omega^2 (0)= m_f^2$. It is clear from
$(\ref{dr1})$ that if $k=0$ then $\omega^2 (0) = M_F^2$ and thereby
we can identify the fermion effective mass at zero temperature as
the rest mass of the fermion, i. e. $m_f^2=M_F^2$. So the gauge
invariant fermion mass, which is generated by the SU(N) gauge
interaction of the massless fermion with the vacuum, is: \be m_f^2 =
\frac{g^2 C(R)}{8} \frac{\mu_f^2}{\pi^2}, \la{mas} \ee being $\mu_f$
a free parameter.

\subsection{Polarization tensor and gauge boson mass}

The gauge boson mass is due to the charge fluctuations of the
vacuum. This mass is calculated following the general procedure
presented in the introduction. The most general form of the
polarization tensor which preserves invariance under rotations,
translations and gauge transformations is \cite{weldon2}: \be
\Pi_{\mu \nu}(K) = P_{\mu \nu} \Pi_T (K) + Q_{\mu \nu} \Pi_L (K),
\la{pot} \ee where the Lorentz-invariant functions $\Pi_L$ and
$\Pi_T$, which characterize the longitudinal and transverse modes
respectively, are obtained by contraction: \bea
\Pi_L(K)=-\frac{K^2}{k^2}u^{\mu}u^{\nu}\Pi_{\mu \nu}, \la{potl} \\
\Pi_T(K)=-\frac{1}{2}\Pi_L + \frac{1}{2}g^{\mu \nu} \Pi_{\mu \nu}.
\la{pott}
\eea

The bosonic dispersion relations are obtained by looking at the
poles of the full propagator, which results of adding all vacuum
polarization insertions. The full bosonic propagator is
\cite{weldon2}: \be D_{\mu \nu}(K) = \frac{Q_{\mu \nu}}{K^2 - \Pi_L
(K)} + \frac{P_{\mu \nu}}{K^2 - \Pi_T (K)} - (\xi - 1) \frac{K_\mu
K_\nu}{K^4} , \la{fbp} \ee where $\xi$ is a gauge parameter. The
gauge invariant dispersion relations, describing the two propagation
modes, are found for: \bea
K^2 - \Pi_L(K) = 0, \la{bdrl} \\
K^2 - \Pi_T(K) = 0. \la{bdrt}
\eea

The one-loop contribution to the vacuum polarization from the
diagram, which can be constructed in this theory, is given by \be
\Pi_{\mu \nu} (K)= i g^2 C(R) \int \frac{d^4p}{(2\pi )^4} Tr
\left[ {\gamma }_\mu S(p) {\gamma }_\nu S(p+K) \right],  \la{ptdf}
\ee where $S$ is the fermion propagator $(\ref{fp})$. Substituting
$(\ref{fp})$ into $(\ref{ptdf})$ the polarization tensor can be
written as $\Pi_{\mu \nu}(K)= \Pi_{\mu \nu}(0)+ \Pi'_{\mu
\nu}(K)$, where $\Pi_{\mu \nu}(0)$ is the zero-density and
zero-temperature contribution and $\Pi'_{\mu \nu}(K)$ is the
finite-density and finite-temperature contribution.

It is easy to see that the real part of the finite-density and
finite-temperature contribution $\mbox{Re}\,\Pi'_{\mu \nu}(K)$ is
given by \be \mbox{Re}\,\Pi'_{\mu \nu}(K) = \frac{g^2 C(R)}{2} \int
\frac{d^4p}{\pi^4} \frac{(p^2 + p \cdot K) g^{\mu \nu} - 2p^{\mu}
p^{\nu} - p^{\mu}K^{\nu} - p^{\nu}K^{\mu}}{(p+K)^2} \Gamma_f(p).
\la{rppt} \ee Substituting $(\ref{rppt})$ in $(\ref{potl})$ and
$(\ref{pott})$ and keeping the leading terms in temperature and
chemical potential, we obtain for the high density approximation
($\mu_f \gg k$ and $\mu_f \gg \omega$) that: \bea
\mbox{Re}\,\Pi'_L(K)=3 M_B^2 \left[ 1 - \frac{\omega}{2k} \log
\frac{\omega+k}{\omega-k} \right], \la{rplo} \\
\mbox{Re}\,\Pi'_T(K)=\frac{3}{2} M_B^2 \left[ \frac{\omega^2}{k^2} +
\left( 1 - \frac{\omega^2}{k^2} \right) \frac{\omega}{2k} \log
\frac{\omega+k}{\omega-k} \right], \la{rptr} \eea where the gauge
boson effective mass $M_B$ is: \be M_B^2(T, \mu_f)= \frac{1}{6} N
g^2 T^2 + \frac{1}{2} g^2 C(R) \left[ \frac{T^2}{6} +
\frac{\mu_f^2}{2 \pi^2} \right], \la{effnab} \ee being $N$ the gauge
group dimension. The non-abelian effective mass $(\ref{effnab})$ is
in agreement with \cite{lebellac}. The abelian gauge boson,
associated with a U(1) gauge invariant theory, acquires an effective
mass $M_{B(a)}$ defined by \be M_{B(a)}^2(T, \mu_f)= e^2 \left[
\frac{T^2}{6} + \frac{\mu_f^2}{2 \pi^2} \right], \la{effab} \ee
being $e$ the interaction coupling constant associated with the U(1)
abelian gauge group. The abelian effective mass $(\ref{effab})$ is
in agreement with \cite{braaten}. Because the vacuum is described by
a virtual medium at $T=0$, then the non-abelian gauge boson
effective mass generated by the quantum fluctuations of the vacuum
is: \be M_{B(na)}^2(0, \mu_f)= M_{B(na)}^2 = g^2 C(R)
\frac{\mu_f^2}{4\pi^2}, \la{bemT0} \ee and the abelian gauge boson
effective mass generated by the quantum fluctuations of the vacuum
is: \be M_{B(a)}^2(0, \mu_f)= M_{B(a)}^2 = e^2
\frac{\mu_f^2}{2\pi^2}, \la{bemT0} \ee in agreement with the result
obtained at finite density and zero temperature \cite{altherr}. For
the limit $k \ll M_{B_\mu}$, it is possible to obtain the dispersion
relations for the transverse and longitudinal propagation modes
\cite{weldon2}: \bea
\omega_L^2=M_{B}^2 + \frac{3}{5}k_L^2 + \ldots\la{disl} \\
\omega_T^2=M_{B}^2 + \frac{6}{5}k_T^2 + \ldots \la{dist} \eea We
note that $(\ref{disl})$ and $(\ref{dist})$ have the same value when
the three-momentum goes to zero. It is clear from $(\ref{disl})$ and
$(\ref{dist})$ that for $k=0$ then $\omega^2(0)=M_{B}^2$ and it is
possible to recognize the gauge boson effective mass as the true
gauge boson mass. The non-abelian gauge boson mass is: \be
m_{b(na)}^2 = M_{B(na)}^2 = g^2 C(R) \frac{\mu_f^2}{4\pi^2},
\la{nabosmas} \ee and the abelian gauge boson mass is:\be m_{b(a)}^2
= M_{B(a)}^2 = e^2 \frac{\mu_f^2}{2\pi^2}. \la{abosmas} \ee We
observe that the gauge boson mass is a function on the chemical
potential that is a free parameter in this effective model. It is
important to note that if the fermionic chemical potential has an
imaginary value, then the gauge boson effective mass, given by
$(\ref{nabosmas})$ or $(\ref{abosmas})$, would be negative
\cite{bluhm}.

\seCtion{SMWHS case}

In this section, following the same procedure as in the previous
one, we calculate the fermions and electroweak gauge bosons masses
for the case in which the dynamics of the physical vacuum is
described by the SMWHS. The physical vacuum is assumed to be a
virtual medium at zero temperature constituted by virtual massless
quarks, antiquarks, leptons and antileptons interacting among
themselves through massless $G$ gluons (for the case of the quarks
and antiquarks), massless $W^{\pm}$ electroweak gauge bosons,
massless $W^3$ gauge bosons and massless $B$ bosons. In this quantum
medium there exist an excess of virtual antifermions over virtual
fermions. This fact is described by non-vanishing chemical
potentials associated with the different fermion flavors. The
chemical potentials are represented for the six quarks by the
symbols $\mu_u, \mu_d, \mu_c, \mu_s, \mu_t, \mu_b$. For the chemical
potentials of the charged leptons we use $\mu_e, \mu_{\mu},
\mu_{\tau}$ and for neutrinos $\mu_{\nu_e},
\mu_{\nu_{\mu}},\mu_{\nu_{\tau}}$. These non-vanishing chemical
potentials are input parameters in the effective model of particle
mass generation.

The dynamics of the vacuum associated with the strong interaction
is described by Quantum Chromodynamics (QCD), while the
electroweak dynamics of the physical vacuum is described by the
$SU(2)_L \times U(1)_Y$ electroweak standard model without Higgs
sector. This model is defined by the following Lagrangian density:
\be {\cal L}_{ew} = {\cal L}_{YM} + {\cal L}_{FB} + {\cal L}_{GF}
+ {\cal L}_{FP}, \la{ldm} \ee where ${\cal L}_{YM}$ is the
Yang-Mills Lagrangian density, ${\cal L}_{FB}$ is the
fermionic-bosonic Lagrangian density, ${\cal L}_{GF}$ is the gauge
fixing Lagrangian density and ${\cal L}_{FP}$ is the Fadeev-Popov
Lagrangian density. The ${\cal L}_{YM}$ is given by \be {\cal
L}_{YM} = -\frac{1}{4} W_{A}^{\mu \nu} W_{\mu \nu}^{A}
-\frac{1}{4} F^{\mu \nu} F_{\mu \nu} \la{lym} \ee where $W_{\mu
\nu}^{A} =\partial_{\mu} W_{\nu}^A -
\partial_{\mu} W_{\mu}^A + g_w F^{ABC} W_{\mu}^B W_{\nu}^C $ is the
energy-momentum tensor associated with the $SU(2)_L$ group and $
F_{\mu \nu} =
\partial_{\mu}B_{\nu} - \partial_{\mu} B_{\mu}$ is the one
associated with the $U(1)_Y$ group. The ${\cal L}_{FB}$ is written
as: \be {\cal L}_{FB} = i \bar{\mbox{L}} \gamma^{\mu}\mbox{D}_\mu
\mbox{L} + i \psi^i_R \gamma^{\mu}\mbox{D}_\mu \psi^i_R + i
\psi^I_R \gamma^{\mu}\mbox{D}_\mu \psi^I_R, \la{lafb} \ee where
$\mbox{D}_\mu \mbox{L} = ( \partial_{\mu} + ig_e Y_L B_\mu/2 +
ig_w T_i W_{\mu}^i ) \mbox{L}$ and $\mbox{D}_\mu \mbox{R} =
(\partial_{\mu} + ig_e Y_R B_\mu/2 ) \mbox{R}$, being $g_w$ the
gauge coupling constant associated with the $SU(2)_L$ group, $g_e$
the one associated to the $U(1)_Y$ group, $Y_L = -1$, $Y_R = -2$
and $T_i = \sigma_i /2$. The $SU(2)_L$ left-handed doublet
$(\mbox{L})$ is given by \be \mbox{L}= {\psi^i \choose {\psi^I}}_L
. \ee

\subsection{Masses of the fermions}

Initially we consider the propagation of massless fermions in a
medium at finite density and finite temperature. The fermion masses
are calculated following the same procedure as in the previous
section. For a non-abelian gauge theory with parity violation and
quirality conservation like the SMWHS, the real part of the
self-energy for a massless fermion is \cite{quimbay}: \be
\mbox{Re}\,\Sigma'(K)=- K{\hspace{-3.1mm}\slash}(a_{L}P_L
+a_{R}P_R)- u{\hspace{-2.2mm}\slash}(b_{L}P_L +b_{R}P_R), \ee where
$P_L \equiv\frac{1}{2}(1-\gamma_5)$ and $P_R
\equiv\frac{1}{2}(1+\gamma_5)$ are respectively the left- and
right-handed chiral projectors. The functions $a_L$, $a_R$, $b_L$
and $b_R$ are the chiral projections of the Lorentz-invariant
functions $a$, $b$ and they are defined in the following way: \bea
a &=& a_L P_L + a_R P_R, \\
b &=& b_L P_L + b_R P_R.
\eea
The inverse fermion propagator is given by
\be
S^{-1}(K)= {\cal L}{\hspace{-2.5mm}\slash} P_L +
\Re{\hspace{-2.5mm}\slash} P_R \la{ifp}
\ee
where:
\bea
{\cal L}^{\mu} &=& ( 1 + a_L) K^{\mu} + b_L u^{\mu} \\
{\Re}^{\mu} &=& ( 1 + a_R) K^{\mu} + b_R u^{\mu} \eea The fermion
propagator follows from the inversion of $(\ref{ifp})$: \bea
S=\frac{1}{D}\left[\left({\cal L}^2\Re{\hspace{-2.5mm}
\slash}\right)P_L + \left(\Re^2{\cal L}{\hspace{-2.5mm}\slash}
\right)P_R \right].\la{p1} \eea being $D(\omega,k)={\cal L}^2
{\Re}^2$. The poles of the propagator correspond to values $\omega$
and $k$ for which the determinant $D$ in (\ref{p1}) vanishes: \be
{\cal L}^2 {\Re}^2 =0.\la{d} \ee In the rest frame of the dense
plasma $u=(1,\vec 0)$, Eq.$(\ref{d})$ leads to the fermion
dispersion relations for a chirally invariant gauge theory with
parity violation, as the case of the SMWHS. Thus, the fermion
dispersion relations are given by \cite{quimbay} \bea \left[ \omega
(1+a_L)+b_L \right]^2- k^2 \left[ 1+a_L \right]^2 &=& 0,
\la{dra}\\
\left[ \omega (1+a_R)+b_R \right]^2-k^2 \left[ 1+a_R \right]^2 &=&
0. \la{drb} \eea Left- and right-handed components of the fermion
dispersion relations obey decoupled relations. The Lorentz invariant
functions $a(\omega,k)$ and $b(\omega,k)$ are calculated from
expressions $(\ref{lifa})$ and $(\ref{lifb})$ through the real part
of the fermion self-energy. This self-energy is obtained adding all
the posibles gauge boson contributions admited by the Feynman rules
of the SMWHS.

We work in the basis of gauge bosons given by $B_\mu$, $W_{\mu}^3$,
$W_{\mu}^{\pm}$, where the charged electroweak gauge bosons are
$W_{\mu}^{\pm} = (W_{\mu}^1 \mp iW_{\mu}^2)/ \sqrt{2}$. The diagrams
with an exchange of $W^{\pm}$ gauge bosons induce a flavor change in
the incoming fermion $i$ to a different outgoing fermion $j$.

\subsubsection{Quark masses}

For the quark sector, in the case of the flavor change contributions
mentioned, the flavor $i$ $(I)$ of the internal quark (inside the
loop) runs over the up $(i)$ or down $(I)$ quarks flavors according
to the type of the external quark (outside the loop). As each
contribution to the quark self-energy is proportional to
$(\ref{Alead})$-$(\ref{Blead})$, the functions $a_L$, $a_R$, $b_L$
and $b_R$ are given by \bea a_L(\omega,k)_{ij} &=&
[f_{S}+f_{W^3}+f_{B}] A(\omega,k,\mu_{i}) + \sum_{I}
f_{W^{\pm}} A(\omega,k,\mu_{I}), \la{al} \\
b_L(\omega,k)_{ij} &=& [f_{S}+f_{W^3}+f_{B}] B(\omega,k,\mu_{i}) +
\sum_{I}
f_{W^{\pm}} B(\omega,k,\mu_{I}), \la{bl} \\
a_R(\omega,k)_{ij} &=&
[f_{S}+f_{B}] A(\omega,k,\mu_{i}), \la{ar} \\
b_R(\omega,k)_{ij} &=&
[f_{S}+f_{B}] B(\omega,k,\mu_{i}). \la{br}
\eea
The coefficients $f$ are:
\bea
f_{S} &=& \frac{4}{3}g_s^2 \delta_{ij}, \la{cfs}\\
f_{W^3}&=& \frac{1}{4} g_w^2 \delta_{ij}, \la{cfw3} \\
f_{B} &=& \frac{1}{4} g_e^2 \delta_{ij}, \la{cfb} \\
f_{W^{\pm}} &=& \frac{1}{2} g_w^2 K_{il}^{+}K_{lj},\la{cfwmm} \eea
where ${\it K}$ represents the CKM matrix and $g_s$ is the strong
running coupling constant associated with the $SU(3)_C$ group. The
integrals $A(\omega,k,\mu_f)$ and $B(\omega,k,\mu_f)$ are obtained
in the high density approximation $(\mu_f \gg k$ and $\mu_f \gg
\omega )$ and, keeping the leading terms in temperature and chemical
potential, they are given by \bea A(\omega,k,\mu_f) &=& \frac{1}{8
k^2} \left( T^2+ \frac{\mu_f^2} {\pi^2}
\right)\left[1-\frac{\omega}{2k} \log \frac{\omega+k} {\omega-k}
\right] \la{ALR},
\\
B(\omega,k,\mu_f) &=& \frac{1}{8 k^2} \left( T^2+ \frac{\mu_f^2}
{\pi^2}\right)\left[\frac{\omega^2-k^2}{2k} \log \frac{\omega+k}
{\omega-k}-\omega \right].\la{BLR} \eea

The chiral projections of the Lorentz-invariant functions are:
\bea a_{L}(\omega,k)_{ij} &=&
\frac{1}{8k^2}\left[1-F(\frac{\omega}{k})
\right]\left[l_{ij}(T^2+\frac{\mu_i^2}{\pi^2})+ h_{ij}(T^2+
\frac{\mu_i^2}{\pi^2})\right], \la{aL} \\
b_{L}(\omega,k)_{ij} &=& -\frac{1}{8k^2}\left[\frac{\omega}{k}+
(\frac{k}{\omega}-\frac{\omega}{k})F(\frac{\omega}{k})\right]\left
[l_{ij}(T^2+\frac{\mu_i^2}{\pi^2})+
h_{ij}(T^2+\frac{\mu_i^2}{\pi^2})
\right], \nn \la{bL} \\
\\
a_{R}(\omega,k)_{ij} &=& \frac{1}{8k^2}\left[1-F(\frac{\omega}{k})
\right]\left[r_{ij}(T^2+\frac{\mu_i^2}{\pi^2}) \right], \la{aR} \\
b_{R}(\omega,k)_{ij} &=& -\frac{1}{8k^2}\left[\frac{\omega}{k}+
(\frac{k}{\omega}-\frac{\omega}{k})F(\frac{\omega}{k})\right]
\left[r_{ij}(T^2+\frac{\mu_i^2}{\pi^2})\right], \la{bR} \eea where
$F(x)$ is \be F(x)=\frac{x}{2} \log \left(\frac{x+1}{x-1}, \right)
\ee and the coefficients $l_{ij}$, $h_{ij}$ and $r_{ij}$ are given
by \bea l_{ij} &=& \left( \frac{4}{3}g_s^2 + \frac{1}{4}g_w^2 +
\frac{1}{4}g_e^2 \right)\delta_{ij}, \\
h_{ij} &=& \sum_l \left(\frac{g_w^2}{2}\right) K_{il}^+ K_{lj}, \\
r_{ij} &=& \left(\frac{4}{3}g_s^2 + \frac{1}{4}g_e^2 \right)
\delta_{ij}. \eea Substituting $(\ref{aL})$-$(\ref{bL})$ into
$(\ref{dra})$, and $(\ref{aR})$-$(\ref{bR})$ into $(\ref{drb})$,
we obtain for the limit $k \ll M_{(i,I)_{L,R}}$ that: \be
\omega^2(k) = M_{(i,I)_{L,R}}^2 \left[ 1 + \frac{2}{3}
\frac{k}{M_{(i,I)_{L,R}}} + \frac{5}{9}
\frac{k^2}{M_{(i,I)_{L,R}}^2} + \dots \right], \la{drsmlr} \ee
where \bea M_{(i,I)_L}^2(T,\mu_f) &=& (l_{ij}+h_{ij})
\frac{T^2}{8} + l_{ij} \frac{\mu_{(i,I)_L}^2}{8 \pi^2} + h_{ij}
\frac{\mu_{(I,i)_L}^2}
{8 \pi^2}, \la{emLq}\\
M_{(i,I)_R}^2(T,\mu_f)  &=& r_{ij} \frac{T^2}{8} + r_{ij}
\frac{\mu_{(i,I)_R}^2}{8 \pi^2}. \la{emRq} \eea As it was
explained in the previous section, we are interested in the
effective masses at $T=0$. For this case: \bea
M_{(i,I)_L}^2(0,\mu_f) &=& l_{ij} \frac{\mu_{(i,I)_L}^2}{8 \pi^2}
+
h_{ij} \frac{\mu_{(I,i)_L}^2}{8 \pi^2}, \la{emLT0}\\
M_{(i,I)_R}^2(0,\mu_f)  &=& r_{ij} \frac{\mu_{(i,I)_R}^2}{8 \pi^2}.
\la{emRT0}
\eea

Following the same argument as in section 2, we can identify the
quark effective masses at zero temperature with the rest masses of
quarks. Coming from the left-handed and right-handed
representations, we find that the masses of the left-handed quarks
are: \bea m_{i}^2 = \left[ \frac{4}{3}g_s^2 + \frac{1}{4}g_w^2 +
\frac{1}{4} g_e^2 \right] \frac{\mu_{i_L}^2}{8 \pi^2} + \left[
\frac{1}{2}g_w^2
\right] \frac{\mu_{I_L}^2}{8 \pi^2}, \la{uqpl} \\
m_{I}^2 = \left[ \frac{4}{3}g_s^2 + \frac{1}{4}g_w^2 + \frac{1}{4}
g_e^2 \right] \frac{\mu_{I_L}^2}{8 \pi^2} + \left[
\frac{1}{2}g_w^2 \right] \frac{\mu_{i_L}^2}{8 \pi^2},  \la{dqpl}
\eea and the masses of the right-handed quarks are: \bea m_{i}^2 =
\left[ \frac{4}{3}g_s^2 + \frac{1}{4} g_e^2 \right]
\frac{\mu_{i_R}^2}{8 \pi^2}, \la{uqpr} \\
m_{I}^2 = \left[ \frac{4}{3}g_s^2 + \frac{1}{4} g_e^2 \right]
\frac{\mu_{I_R}^2}{8 \pi^2}, \la{dqpr} \eea where the couple of
indexes $(i,I)$ run over quarks $(u, d)$, $(c, s)$ and $(t, b)$.
It is known that the masses of the left-handed quarks are the same
that the ones of right-handed quarks. This means that the
left-handed quark chemical potentials $\mu_{q_L}$ are different
than the right-handed quark chemical potentials $\mu_{f_R}$.

If we call \bea a_q &=&\frac{1}{8 \pi^2} \left[ \frac{4}{3}g_s^2 +
\frac{1}{4}g_w^2 +\frac{1}{4} g_e^2 \right], \la{caq}\\
b_q &=&\frac{1}{8 \pi^2} \left[ \frac{1}{2}g_w^2 \right], \la{cbq}
\eea it is easy to see that the expressions $(\ref{uqpl})$ and
$(\ref{dqpl})$ lead to \bea
\mu_{u_L}^2 &=& \frac{a_q m_u^2 - b_q m_d^2}{a_q^2 - b_q^2}, \la{pqu2} \\
\mu_{d_L}^2 &=& \frac{-b_q m_u^2 + a_q m_d^2}{a_q^2 - b_q^2},
\la{pqd2} \eea and if we call \be c_q =\frac{1}{8 \pi^2} \left[
\frac{4}{3}g_s^2 + \frac{1}{4} g_e^2 \right], \la{ccq} \ee the
expressions $(\ref{uqpr})$ and $(\ref{dqpr})$ can be written as
\bea
\mu_{u_R}^2 &=& \frac{m_u^2}{c_q}, \la{pquR2} \\
\mu_{d_R}^2 &=& \frac{m_u^2}{c_q}, \la{pqdR2} \eea and similar
expressions for the other two quark doublets $(c, s)$ and $(t,
b)$.

If we take the experimental central values for the strong constant
$\alpha_s=0.1176$, the fine-structure constant as
$\alpha_e=7.2973525376 \times 10^{-3}$ and the cosine of the
electroweak mixing angle as $\cos \theta_w =
M_W/M_Z=80.403/91.1876=0.881732$ \cite{pdg}, then we have that
$g_s=1.21565$, $g_w=0.641911$ and $g_e=0.34344$. Putting the
central values for the masses of the quarks, given by \cite{pdg}
$m_u = 0.00225$ GeV, $m_d = 0.005$ GeV, $m_c = 1.25$ GeV, $m_s =
0.095$ GeV, $m_t = 172.371$ GeV and $m_b = 4.20$ GeV,  into the
expressions $(\ref{pqu2})$, $(\ref{pqd2})$, $(\ref{pquR2})$ and
$(\ref{pqdR2})$, we obtain that the squares of the left-handed
quark chemical potentials are given by \bea \mu_{u_L}^2 &=& 9.9068
\times
10^{-5}, \la{vpqu2} \\ \mu_{d_L}^2 &=& 9.2896 \times 10^{-4},\\
\mu_{c_L}^2 &=& 59.2015, \\
\mu_{s_L}^2 &=& -5.4612, \\ \mu_{t_L}^2 &=& 1.1263 \times 10^{6}, \\
\mu_{b_L}^2 &=& -1.0969 \times 10^{5}, \la{vpqb2}\eea and the the
squares of the right-handed quark chemical potentials are \bea
\mu_{u_R}^2 &=& 1.9987 \times
10^{-4}, \la{vpqu2} \\ \mu_{d_R}^2 &=& 9.8701 \times 10^{-4},\\
\mu_{c_R}^2 &=& 61.6883, \\
\mu_{s_R}^2 &=& 0.3563, \\ \mu_{t_R}^2 &=& 1.1730 \times 10^{6}, \\
\mu_{b_R}^2 &=& 696.436, \la{vpqb2}\eea where the left- and
right-handed chemical potentials are given in GeV$^2$ units.

\subsubsection{Lepton masses}

For the lepton sector, the contributions to the fermion
self-energy are proportional to $(\ref{Alead})$-$(\ref{Blead})$
and the functions $a_L$, $a_R$, $b_L$ and $b_R$ are given by \bea
a_L(\omega,k)_{ij} &=& [f_{W^3}+f_{B}] A(\omega,k,\mu_{i}) +
\sum_{I}
f_{W^{\pm}} A(\omega,k,\mu_{I}), \la{al} \\
b_L(\omega,k)_{ij} &=& [f_{W^3}+f_{B}] B(\omega,k,\mu_{i}) +
\sum_{I}
f_{W^{\pm}} B(\omega,k,\mu_{I}), \la{bl} \\
a_R(\omega,k)_{ij} &=& [f_{B}] A(\omega,k,\mu_{i}), \la{ar} \\
b_R(\omega,k)_{ij} &=& [f_{B}] B(\omega,k,\mu_{i}), \la{br} \eea
being for this case $f_{W^{\pm}}= g_w^2/2$ and the other
coefficients $f_{W^3}$ and $f_{B}$ are given by $(\ref{cfw3})$ and
$(\ref{cfb})$, respectively.

The dispersion relation for the leptons are similar to the
relations $(\ref{drsmlr})$, but in this case the effective masses
$(\ref{emLq})$  and $(\ref{emRq})$ are given by \bea
M_{(i,I)_L}^2(T,\mu_f) &=& (l+h) \frac{T^2}{8} + l
\frac{\mu_{(i,I)_L}^2}{8 \pi^2} + h \frac{\mu_{(I,i)_L}^2}
{8 \pi^2}, \la{emLl}\\
M_{(i,I)_R}^2(T,\mu_f)  &=& r \frac{T^2}{8} + r
\frac{\mu_{(i,I)_R}^2}{8 \pi^2}, \la{emRl} \eea where the
coefficients $l$, $h$ and $r$, for the charged leptons are given
by \bea
l &=& \left( \frac{1}{4}g_w^2 + \frac{1}{4}g_e^2 \right), \\
h &=& \left( \frac{1}{2} g_w^2 \right), \\
r &=& \left(\frac{1}{4}g_e^2 \right), \eea and for the neutrinos
these coefficients are: \bea
l &=& \left( \frac{1}{4}g_w^2 \right), \\
h &=& \left( \frac{1}{2} g_w^2 \right), \\
r &=& 0. \eea

The leptonic effective masses $(\ref{emLl})$  and $(\ref{emRl})$
at zero temperature can be interpreted as the masses of the
leptons. Coming from the left-handed and right-handed
representations, respectively, we find that the masses of the
left-handed leptons are given by: \bea m_{i}^2 = \left[
\frac{1}{4}g_w^2 \right] \frac{\mu_{i_L}^2}{8 \pi^2} + \left[
\frac{1}{2}g_w^2
\right] \frac{\mu_{I_L}^2}{8 \pi^2}, \la{nlpl} \\
m_{I}^2 = \left[ \frac{1}{4}g_w^2 + \frac{1}{4} g_e^2 \right]
\frac{\mu_{I_L}^2}{8 \pi^2} + \left[ \frac{1}{2}g_w^2 \right]
\frac{\mu_{i_L}^2}{8 \pi^2},  \la{elpl} \eea and the masses of the
right-handed charged leptons are \bea m_{I}^2 = \left[ \frac{1}{4}
g_e^2 \right] \frac{\mu_{I_R}^2}{8 \pi^2}, \la{elpr} \eea where
the couple of indexes $(i,I)$ run over leptons $(\nu_e, e)$,
$(\nu_{\mu}, \mu)$ and $(\nu_{\tau}, \tau)$. We note that our
effective model predicts that neutrinos are massive. The $W^3$ and
$W^{\pm}$ interactions among the massless neutrinos with the
physical vacuum are the origin of the left-handed neutrinos
masses, as we can observe from $(\ref{nlpl})$.

If we call \bea a_l&=&\frac{1}{8 \pi^2} \left[ \frac{1}{4}g_w^2
\right], \la{cal}\\
b_l &=&\frac{1}{8 \pi^2} \left[ \frac{1}{2}g_w^2 \right], \la{cbl} \\
c_l &=&\frac{1}{8 \pi^2} \left[ \frac{1}{4}g_w^2+\frac{1}{4}g_e^2
\right], \la{cbl} \eea then the expressions $(\ref{nlpl})$ and
$(\ref{elpl})$ lead to \bea
\mu_{\nu_L}^2 &=& \frac{c_l m_\nu^2 - b_l m_e^2}{a_l c_l - b_l^2},
\la{pqn2} \\
\mu_{e_L}^2 &=& \frac{-b_l m_\nu^2 + a_l m_e^2}{a_l c_l - b_l^2},
\la{pqe2} \eea and if we call \be d_l =\frac{1}{8 \pi^2} \left[
\frac{1}{4} g_e^2 \right], \la{ccq} \ee the expression
$(\ref{elpr})$ can be written as \be \mu_{e_R}^2 =
\frac{m_e^2}{d_l}, \la{pqeR2} \ee and similar expressions for the
other two lepton doublets $(\nu_\mu, \mu)$ and $(\nu_\tau, \tau)$.
Assuming the neutrinos to be massless, $m_{\nu_e} = m_{\nu_\mu}
=m_{\nu_\tau} =0$, and putting the experimental central values for
the masses of the charged leptons, given by \cite{pdg} $m_e =
0.51099892 \times 10^{-3}$ GeV, $m_\mu= 0.105658369$ GeV, $m_\tau
= 1.77699$ GeV, in the expressions $(\ref{pqn2})$, $(\ref{pqe2})$
and $(\ref{pqeR2})$, we obtain that the squares of the left-handed
lepton chemical potentials are \bea \mu_{\nu_{e_L}}^2 &=& 3.9182
\times 10^{-4}, \la{vpqn2}
\\ \mu_{e_L}^2 &=& -3.0462 \times
10^{-4}, \\ \mu_{\nu_{\mu_L}}^2 &=& 16.8179, \\ \mu_{\mu_L}^2 &=&
-13.0751, \\
\mu_{\nu_{\tau_L}}^2 &=& 4.7569 \times 10^{3}\\ \mu_{\tau_L}^2 &=&
-3.6982 \times 10^{3}, \la{vpqtau2} \eea and the the squares of
the right-handed charged lepton chemical potentials are \bea
\mu_{e_R}^2 &=& 6.9645 \times
10^{-4}, \la{vpqe2} \\ \mu_{\mu_R}^2 &=& 29.8929, \\
\mu_{\tau_R}^2 &=& 8.4551 \times 10^{3} \la{vpqtau2}\eea where the
left- and right-handed chemical potentials are given in GeV$^2$
units. Neutrino masses are not known but direct experimental
results shown that neutrino masses are of order $1$ eV \cite{pdg},
and cosmological interpretations from five-year WMAP observations
find a limit on the total mass of massive neutrinos of $ \Sigma
m_\nu < 0.6$ eV ($95\%$ CL) \cite{wmap5year}. These results assure
that the values of the left-handed lepton chemical potentials
obtained taking neutrinos to be massless will change little if we
take the true small neutrinos masses.

We observe that for five of the six fermion doublets the square of
the chemical potential associated to the down fermion of the
doublet has a negative value. This behavior is observed if there
is a large difference between the masses of the two fermions of
the doublet. This means that it behavior is not observed for the
quark doublet which is formed by the up and down quarks, because
these two quarks have masses very close. In this case, the
chemical potentials associated to these two quarks are positives.

From expressions $(\ref{uqpl})$, $(\ref{dqpl})$, $(\ref{nlpl})$
and $(\ref{elpl})$ it can be seen that our effective model does
not predict the fermions masses values because the fermionic
chemicals potentials $\mu_{f_i}$ are free parameters. However, we
have fixed the values of these $\mu_{f_i}$ starting from the known
experimental values for the fermions masses. This limitation of
our effective model is similar to what happens in the MSM with
Higgs mechanism, in the sense that the masses of the fermions
depend on the Yukawa coupling constants which are free parameters.
In a similar way to what happens here when we find the values of
the vacuum chemical potentials, the Yukawa coupling constants can
be fixed by means of the experimental values of the fermion
masses.

\subsection{Masses of the electroweak gauge bosons}

The masses of the electroweak gauge bosons are originated from the
charge fluctuations of the vacuum. These masses are calculated
following a sequence of steps that we present now: In the first
step we write the one-loop polarization tensor at finite density
and finite temperature; then we calculate the one-loop bosonic
dispersion relations in the high density approximation by
obtention of the poles of the gauge boson propagators; starting
from these dispersion relations we obtain the electroweak gauge
boson effective masses at finite density and finite temperature;
finally we identify these effective masses at zero temperature
with the masses of the electroweak gauge bosons.

To evaluate the bosonic polarization tensor associated with the
$W^{\pm}_\mu$, $W^3_\mu$, $B_\mu$ gauge boson propagators, we
follow the same procedure as in the section 2.2. On the other
hand, applying the expressions $(\ref{nabosmas})$ and
$(\ref{abosmas})$ in the SMWHS, we obtain that the masses of the
gauge bosons are: \bea M_{W^\pm}^2 &=& \frac{g_w^2}{2} \,\
\frac{\mu_{u_L}^2 + \mu_{d_L}^2 +\mu_{c_L}^2 -
\mu_{s_L}^2+\mu_{t_L}^2 - \mu_{b_L}^2+ \sum_{i=1}^{3}
(\mu_{\nu_{i_L}}^2 -
\mu_{e_{i_L}}^2 )}{4\pi^2}, \la{mw+-} \\
M_{W^3}^2 &=& \frac{g_w^2}{4} \,\ \frac{\mu_{u_L}^2 + \mu_{d_L}^2
+\mu_{c_L}^2 - \mu_{s_L}^2+\mu_{t_L}^2 - \mu_{b_L}^2+
\sum_{i=1}^{3} (\mu_{\nu_{i_L}}^2 -
\mu_{e_{i_L}}^2 )}{2\pi^2}, \la{mw3} \\
M_{B}^2 &=& \frac{g_e^2}{4} \,\ \frac{\mu_{u_L}^2 + \mu_{d_L}^2
+\mu_{c_L}^2 - \mu_{s_L}^2+\mu_{t_L}^2 - \mu_{b_L}^2+
\sum_{i=1}^{3} (\mu_{\nu_{i_L}}^2 - \mu_{e_{i_L}}^2 )}{2\pi^2},
\la{mb} \eea where the sum runs over the three lepton families. It
is important to remember that if the left-handed fermionic
chemical potential has an imaginary value, then its contribution
to the gauge boson effective mass, as in the case
$(\ref{nabosmas})$ or $(\ref{abosmas})$, would be negative. This
fact means that finally the contribution from each fermionic
chemical potential to the masses of the gauge bosons is always
positive.

Substituting the obtained left-handed fermionic chemicals
potentials values $(\ref{vpqu2})$-$(\ref{vpqb2})$ and
$(\ref{vpqn2})$-$(\ref{vpqtau2})$ into the expressions
$(\ref{mw+-})$-$(\ref{mb})$, we obtain \bea
M_{W^{\pm}}= M_{W^{3}}= 80.403 \,{\mbox GeV}, \la{masw} \\
M_{B}= 42.9954 \,{\mbox GeV}, \la{masb} \eea where the $M_W$ value
computed is in agreement with its experimental value given by
$M_W^{exp}= 80.403 \pm 0.029$ GeV \cite{pdg}.

Due to well known physical reasons $W_{\mu}^3$ and $B_\mu$ gauge
bosons are mixed. After diagonalization of the mass matrix, we get
the physical fields $A_\mu$ and $Z_\mu$ corresponding to the
massless photon and the neutral $Z^0$ boson of mass $M_Z$
respectively, with the relations \cite{weinberg1, pestieau}: \bea
M_Z^2 = M_W^2 + M_B^2 , \la{masz} \\
\cos \theta_w = \frac{M_W}{M_Z} \hspace{3.0mm} ,  \hspace{3.0mm}
\sin \theta_w = \frac{M_B}{M_Z}, \la{mix} \eea where $\theta_w$ is
the weak mixing angle: \bea
Z_{\mu}^0 = B_\mu \sin \theta_w - W_{\mu}^3 \cos \theta_w , \la{zbw} \\
A_{\mu} = B_\mu \cos \theta_w + W_{\mu}^3 \sin \theta_w . \la{abw}
\eea Substituting $(\ref{masw})$ and $(\ref{masb})$ into
$(\ref{masz})$ we obtain that: \be M_{Z}= 91.1876 \,{\mbox GeV},
\la{masZ0} \ee also in agreement with its experimental value given
by $M_Z^{exp}= 91.1876 \pm 0.0021$ GeV \cite{pdg}.

Substituting the expressions for the fermionic chemical potentials
given by $(\ref{pqu2})$, $(\ref{pqd2})$, $(\ref{pqn2})$,
$(\ref{pqe2})$ into the expressions $(\ref{mw+-})$, $(\ref{mw3})$,
$(\ref{mb})$, we then obtain that the masses of the electroweak
gauge bosons $W$ and $Z$ in terms of the masses of the fermions
and the running coupling constants of the strong, weak and
electromagnetic interactions are given by \bea
M_W^2 &=& g_w^2 (A_1 + A_2 +A_3-A_4) , \la{mw2mfrc} \\
M_Z^2 &=& (g_e ^2 + g_w^2) (A_1 + A_2 +A_3-A_4), \la{mz2mfrc} \eea
where the parameters $A_1$, $A_2$, $A_3$ and $A_4$ are \bea
A_1 &=& \frac{m_u^2+m_d^2}{B_1}, \la{A1} \\
A_2 &=& \frac{m_c^2-m_s^2+m_t^2-m_b^2}{B_2}, \la{A2} \\
A_3 &=&
\frac{3(m_e^2+m_\mu^2+m_\tau^2)}{B_3}, \la{A3} \\
A_4 &=& \frac{(3+g_e^2/g_w^2)(m_{\nu_e}^2 +m_{\nu_\mu}^2
+m_{\nu_\tau}^2)}{B_3}, \la{A4} \eea being \bea
B_1 &=& \frac{4}{3}g_s^2 + \frac{3}{4}g_w^2 + \frac{1}{4}g_e^2, \la{B1} \\
B_2 &=& \frac{4}{3}g_s^2 - \frac{1}{4}g_w^2 + \frac{1}{4}g_e^2, \la{B2} \\
B_3 &=& \frac{3}{4}g_w^2 - \frac{1}{4}g_e^2. \la{B3} \eea

If we take the experimental central values for the strong constant
at the $M_Z$ scale as $\alpha_s(M_Z)=0.1176$, the fine-structure
constant as $\alpha_e=7.2973525376 \times 10^{-3}$ and the cosine
of the electroweak mixing angle as $\cos \theta_w =
M_W/M_Z=80.403/91.1876=0.881732$ \cite{pdg}, then $g_s=1.21565$,
$g_w=0.641911$ and $g_e=0.34344$. Putting these values for $g_s$,
$g_w$ and $g_e$ and the central values for the masses of the
electrically charged fermions, given by \cite{pdg} $m_u = 0.00225$
GeV, $m_d = 0.005$ GeV, $m_c = 1.25$ GeV, $m_s = 0.095$ GeV, $m_t
= 172.371$ GeV, $m_b = 4.20$ GeV, $m_e = 0.51099892 \times
10^{-3}$ GeV, $m_\mu= 0.105658369$ GeV, $m_\tau = 1.77699$ GeV,
into the expressions $(\ref{mw2mfrc})$ and $(\ref{mz2mfrc})$, and
assuming the neutrinos to be massless, $m_{\nu_e} = m_{\nu_\mu}
=m_{\nu_\tau} =0$, we obtain that the theoretical masses of the
$W$ and $Z$ electroweak gauge bosons are \bea M_{W^{\pm}} &=&
80.403 \,{\mbox GeV} \\ M_{Z} &=& 91.1876 \,{\mbox GeV}.
\la{masZ0} \la{masw} \eea These theoretical masses are in
agreement with theirs experimental values given by $M_W^{exp}=
80.403 \pm 0.029$ GeV and $M_Z^{exp}= 91.1876 \pm 0.0021$ GeV
\cite{pdg}. The parameters $A_1$, $A_2$, $A_3$ and $A_3$ in the
expressions $(\ref{mw2mfrc})$ and $(\ref{mz2mfrc})$ have the
following values $A_1=1.30201 \times 10^{-5}$, $A_2=15655$,
$A_3=34.0068$ and $A_4=0$. We observe that $A_2$ is very large
respect to $A_3$ and $A_1$. Through the definition of the
parameter $A_2$, given by $(\ref{A2})$, it is possible to conclude
that the masses of the electroweak gauge bosons come mainly from
the top quark chemical potential $\mu_{t_L}$ and the strong
running coupling constant $g_s$.

\seCtion{Conclusions}

\hspace{3.0mm}

We presented an effective model for particle mass generation in
which we extracted some generic features of the Higgs mechanism
that do not depend on its interpretation in terms of a Higgs
field. The physical vacuum has been assumed to be a medium at zero
temperature which is formed by virtual massless fermions and
antifermions interacting among themselves by means of massless
gauge bosons. The fundamental effective model describing the
dynamics of this physical vacuum is the SMWHS. We have assumed
that each fermion flavor in the physical vacuum has associated
with it a chemical potential $\mu_{f}$ in such a way that there is
an excess of virtual antifermions over virtual fermions. This fact
implies that the vacuum is thought to be a virtual medium having a
net antimatter finite density.

Fermion masses are calculated starting from the fermion
self-energy, which represents the fundamental interactions of a
massless fermion with the physical vacuum. The gauge boson masses
are calculated from the charge fluctuations of the physical
vacuum, which are described by the vacuum polarization tensor. We
have used the following general procedure to calculate these
particle masses: Initially we write the one-loop self-energies and
polarization tensors at finite density and finite temperature;
next we calculate the dispersion relations by obtention of the
poles of the fermion and gauge boson propagators; starting from
these dispersion relations we obtain the fermion and gauge boson
effective masses at finite density and finite temperature; finally
we identify these particle effective masses at zero temperature
with the physical particle masses. This identification can be
performed because in our effective model the virtual medium, at
finite density and zero temperature, represents the physical
vacuum.

Using this effective model for particle mass generation, we obtain
the masses of the electroweak gauge bosons in agreement with their
experimental values. A further result of our effective model is
that the left-handed neutrinos are massive because they have weak
charge.

\hspace{3.0mm}

\section*{Acknowledgments} This work was supported by COLCIENCIAS
(Colombia) under research grant 1101-05-13610, CT215-2003. C. Q.
thank Alfonso Rueda, Carlos Avila, Miguel Angel Vazquez-Mozo, Rafael
Hurtado and Antonio S\'anchez for stimulating discussions.


\end{document}